\documentclass[%
 reprint,
 amsmath,amssymb,
 aps,
]{revtex4-2}

\usepackage{graphicx}
\usepackage{dcolumn}
\usepackage{bm}

\begin{document}

\preprint{APS/123-QED}

\title{Theory of light propagation in arbitrary two-dimensional curved space}

\author{Chenni Xu}
\author{Li-Gang Wang}%
\email{lgwang@zju.edu.cn}
\affiliation{Zhejiang Province Key Laboratory of Quantum Technology and Device, Department of Physics, Zhejiang University, Hangzhou, 310027, Zhejiang, China }%


\date{July 13, 2021}

\begin{abstract}
As an analog model of general relativity, optics on some two-dimensional (2D) curved surfaces has been increasingly paid attention to in the past decade. Here, in light of Huygens-Fresnel principle, we propose a theoretical frame to study light propagation along arbitrary geodesics on any 2D curved surfaces. This theory not only enables us to solve the enigma of ``infinite intensity'' existed previously at artificial singularities on surfaces of revolution, but also
makes it possible to study light propagation on arbitrary 2D curved surfaces. Based on this theory, we investigate the effects of light propagation on a typical surface of revolution, Flamm's paraboloid, as an example, from which one can understand the behavior of light in the curved geometry of Schwarzschild black holes. Our theory provides a convenient and powerful tool for investigations of radiation in curved space.
\end{abstract}

\maketitle


\section{INTRODUCTION}

In general relativity (GR), spacetime is distorted in the vicinity of massive celestial bodies. Dynamics of electromagnetic (EM) waves in the context of strong gravitational fields has attracted increasing attention, ranging from wave optics \cite{Nambu2019}, gravitational lensing \cite{Lupsasca2020}, scattering theory \cite{scattering2009}, as well as photon rings \cite{Johnson2020} which are predicted to ensconce in the shadow of the M87* black hole image recently published by EHT Collaboration \cite{ETH2019}. Despite the flourishing astrophysical explorations, investigations from the perspective of optics are still rare. Because of feeble gravitational effects, measurements and verification of GR phenomena are difficult to perform unless in an astronomical scale. Therefore, researchers have proposed various analogue models to study GR phenomena by table-top equipments in laboratory \cite{Faccio2013}, such as observation of spontaneous Hawking radiation in a flowing Bose-Einstein condensate \cite{Steinhauer2019}, emulation of Schwarzschild precession with a gradient index lens \cite{Chen2021}, mimicking gravitational lensing by a microstructured optical waveguide \cite{Sheng2013}. Another analogue model is to abandon one spatial dimension and fix time coordinate of the four-dimensional (4D) curved spacetime. In this manner, the remaining 2D spatial metric tensor can be depicted as a 2D curved surface embedded in 3D space, and the interplay between EM waves and spatial curvature can be revealed by investigating light propagation on such appropriately fabricated surfaces. Ever since this notion was put forward by Batz and Peschel \cite{Batz2008} in 2008, various optical phenomena have been reported both theoretically \cite{Batz2008, Batz2010, Bekenstein2014, Lustig2017, XuPRA, XuOE, Wang2018, XuNJP, Shao2021} and experimentally \cite{Schultheiss2010, Schultheiss2015, Patsyk2018, Bekenstein2017, Li2018}. Besides optics and photonics, similar studies on curved surfaces have also been extended to surface plasmon polaritons \cite{Arie2019}, acoustic topological insulators \cite{Jing2021} and quantum particles \cite{Longhi2009}. 

The theory of light propagation in 2D curved space was initiated by Batz et al. \cite{Batz2008},
by obtaining a nonlinear Schr\"{o}dinger equation on surfaces of revolution (SORs) with constant Gaussian curvature. Owing to the rotational symmetry of SORs, the curvilinear coordinates on surfaces are conveniently taken along longitudes and latitudes. This paradigm ingeniously simplifies the calculation to a great extent. However, the solution applies exclusively to propagation along longitudinal direction, which is a special one among innumerable geodesics. Indeed, considering light propagation along non-longitudinal directions is more challenging, not only because of the tedious calculation of analytically solving the convariant wave equation, but also the ambiguous physical images that are beyond intuitive imagination. Due to the rotational symmetry of SORs, a light beam launched tangent to a longitude recognizes an axisymmetric distribution of spatial curvature, which guarantees its propagation right along the very longitude. However, such axisymmetry doesn't hold true for light beams with other initial directions, whose trajectories will therefore be bent somehow. Intriguing questions naturally arise, for instance, which pathway would the light beam take and how would the curvature of surface affect its divergence? Besides, it has also been revealed in prior studies \cite{Batz2008, Schultheiss2015, XuOE, XuNJP} that the existed method for calculating light fields on closed SORs collapsed at artificial singularities (such as both the north and south poles on spherical surfaces), leading to an artificial
``infinite intensity'' thereat.

In this paper, we propose an alternative approach to study light propagation along arbitrary geodesics on any curved surfaces, in light of Huygens-Fresnel principle. We assume the light wave propagates along geodesic, the natural path of its ray counterpart in curved space. This approach not only figures out the problems mentioned above, but also in a manner that refrains us from complicated mathematics. Using this approach, we take a Flamm's paraboloid, which is the 2D correspondence of Schwarzschild metric, as an example, and demonstrate the behaviors of both collimated and highly divergent light in such curved space. At last, we figure out the remaining enigma of artificial singularities in the previous method, and suggest some possible schemes for experimental verification. 

\section{RESULTS AND DISCUSSION}
\subsection{Basic theory}
\begin{figure}[btp]
\centering
\includegraphics[width=8.5cm]{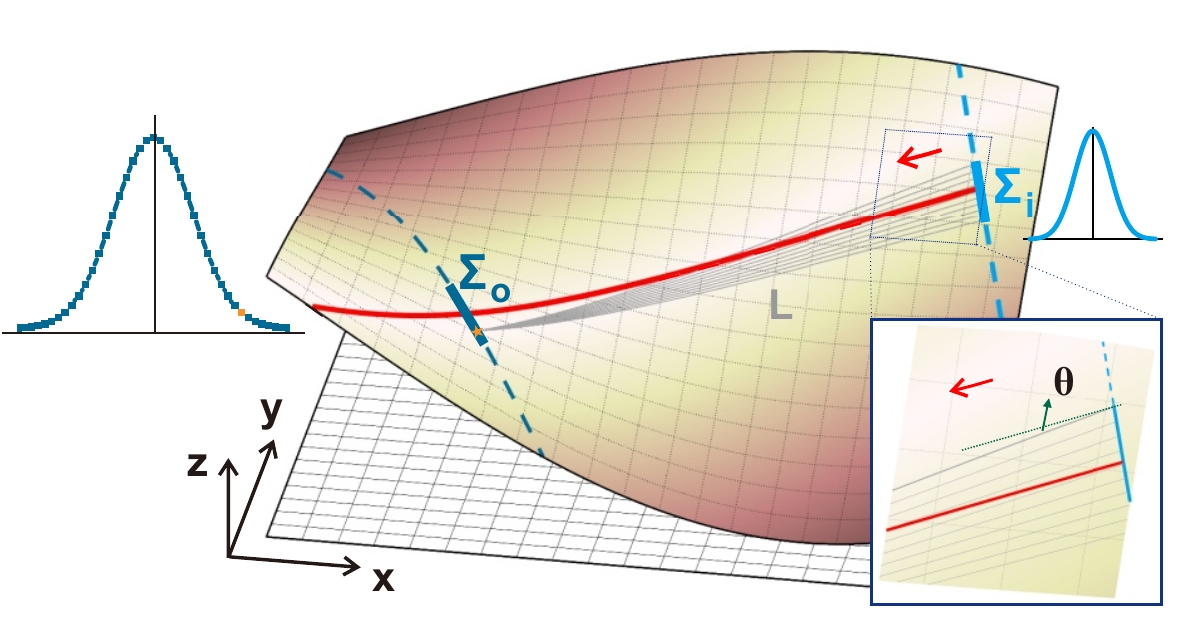}
\caption{Schematic of a 2D curved surface generated from the planar Cartesian coordinates $(x,y)$, here as an example with the height of the surface $z=H(x,y)=\sin{x}\cos{y}$. The red solid line denotes an arbitrary geodesic as the propagation axis of a light beam on this curved surface. $\Sigma _{i}$ and $\Sigma _{o}$ are the two geodesics locally vertical to the propagation axis as the input and output interfaces, respectively. The gray-dark lines are the shortest geodesics from the points on $\Sigma _{i}$ to the orange-color point on $\Sigma _{o}$ and one of the angles between these shortest geodesics and the propagation axis is denoted by $\theta$ in the inset figure.
}%
\label{figure1}%
\end{figure}

Consider an arbitrary 2D curved surface that can be fabricated by deforming a
plane. The points on an arbitrary curved surface can be expressed by the 3D Cartesian coordinates as $ \left[x,y,z=H(x,y)\right]  $, with $x,y$ being the planar Cartesian coordinates and an
arbitrary function $H$ marking the height difference between the curved surface and the $x-y$ plane, as is sketched in Fig. \ref{figure1}. The
corresponding metric of the curved surface is%
\begin{align}
ds^{2} = & \,g_{ij}dx^{i}dx^{j}\nonumber\\
= & \left[  1+\left(  \frac{\partial H}{\partial x}\right) ^{2}\right]
dx^{2}+\left[  1+\left(  \frac{\partial H}{\partial y}\right)  ^{2}\right]
dy^{2} \nonumber \\
& +2\frac{\partial H}{\partial x}\frac{\partial H}{\partial y}dxdy,
\label{metric}%
\end{align}
where the third term indicates that the coordinate system we choose to parametrize the
surface is not orthogonal on the curved surface. On curved surfaces, light rays propagate along the so-called
geodesics, which are the counterparts of the straight lines in flat space.
The geodesic equation is given by $\frac{d^{2}x^{\sigma}}{ds^{2}}+\Gamma_{\mu\nu
}^{\sigma}\frac{dx^{\mu}}{ds}\frac{dx^{\nu}}{ds}=0$, where $\Gamma_{\mu\nu
}^{\sigma}=\frac{1}{2}g^{\sigma\rho}\left(  \frac{\partial g_{\rho\mu}%
}{\partial x^{\nu}}+\frac{\partial g_{\rho\nu}}{\partial x^{\mu}}%
-\frac{\partial g_{\mu\nu}}{\partial x^{\rho}}\right)  $ are Christoffel
connections, $g^{\sigma\rho}$ are the elements of the inverse of the metric tensor
$\mathbf{g}$, and Einstein summation convention is applied, with $x^{\sigma
},x^{\mu},x^{\nu}$ running through $x,y$. Therefore, one can track a
light ray on a curved surface by solving the following equations%
\begin{equation}
\frac{d^{2}x}{ds^{2}}+\Gamma_{xx}^{x}\left(  \frac{dx}{ds}\right)  ^{2}%
+\Gamma_{yy}^{x}\left(  \frac{dy}{ds}\right)  ^{2}+2\Gamma_{xy}^{x}\frac
{dx}{ds}\frac{dy}{ds}=0, \label{geodesic1}%
\end{equation}%
\begin{equation}
\frac{d^{2}y}{ds^{2}}+\Gamma_{xx}^{y}\left(  \frac{dx}{ds}\right)  ^{2}%
+\Gamma_{yy}^{y}\left(  \frac{dy}{ds}\right)  ^{2}+2\Gamma_{xy}^{y}\frac
{dx}{ds}\frac{dy}{ds}=0. \label{geodesic2}%
\end{equation}
Generally, it is difficult to acquire the analytical solution to this equation set, unless some extra properties of surfaces, such as symmetries, are present.
Usually numerical methods, such as the Runge-Kutta method, can be utilized to solve the above equations as long as the step length meets the accuracy requirements.

\begin{figure*}[btp]
\centering
\includegraphics[width=18cm]{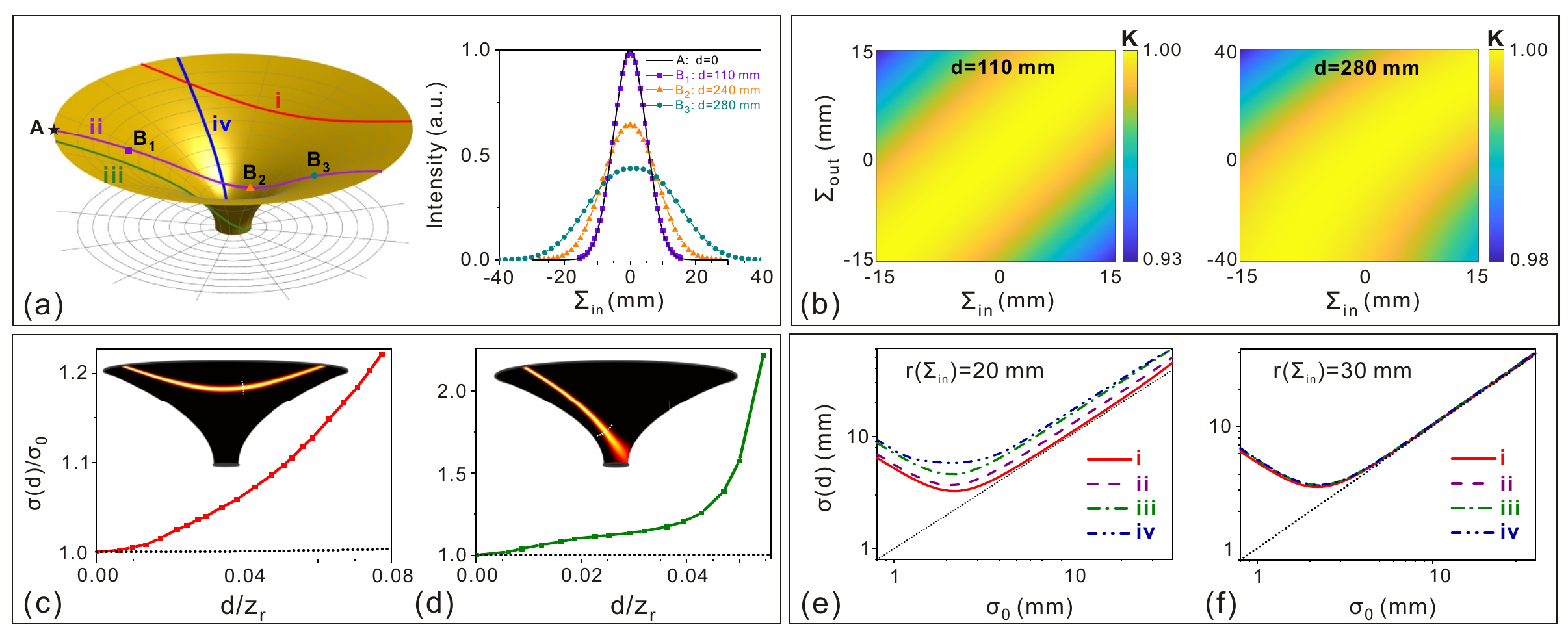}
\caption{(a) (Left) Sketch of a Flamm's paraboloid, and (Right) the output-field intensity distribution at different propagation distance $d$ for the cases of a well-collimated Gaussian beam launched from Point A (the incident end) to Points B$_{1}$, B$_{2}$, and B$_{3}$ (different output ends) along the curve (ii). The curves (i)-(iv) with different colors on the surface show four typical geodesics along which a light beam propagates. (b) Magnitudes of the obliquity factor $K$ between the input and output ends under different values of $d=110$mm and $280$mm. (c)(d) Changes of the beam width $\sigma(d)$ of light along the geodesics (i) and (iii) in (a), respectively. The insets in (c) and (d) show the intensity evolution along the geodesics (i) and (iii), respectively, and the near-horizontal dot lines in (c) and (d) denote the changes of $\sigma(d)$ in flat space. (e)(f) The dependence of the output $\sigma(d)$ on the initial $\sigma_{0}$ along the four geodesics in (a) with the fixed value of $d=220$mm, when the input end is locating differently at (e) $r=200$ mm and (f) $r=300$ mm. Other parameters are $r_{s}=20$ mm, $\lambda=7\times10^{-5}$ m, $r(\Sigma_{i})=200$ mm, $\sigma_{0}=10$ mm.}%
\label{figure2}%
\end{figure*}

Now let us consider the propagation of light on curved surfaces.
First, we build up the coordinates on a curved surface, as is shown in Fig. \ref{figure1}. The optical propagation axis of a light beam is taken along the arbitrary geodesic that we are interested in, for example, the red line on the surface. Vertical to the optical propagation axis, the input and output interfaces are taken along the orthogonal geodesics, respectively, denoted by $\Sigma _{i}$ and $\Sigma _{o}$ on that surface. In light of Huygens-Fresnel principle, each point $S_{i}$ on the input interface $\Sigma _{i}$ is a source
of secondary spherical wavelet. The secondary wavelets emanating from
different points on the initial interface interfere mutually, whose
superposition forms the far-field wavefront at $P_{o}$ on the output interface $\Sigma _{o}$. Put it mathematically,
the complex amplitude at $P_{o}$ on $\Sigma _{o}$ is described as
\begin{equation}
\Phi_{o}(P_{o})=\sqrt{\frac{1}{i\lambda}}\int\limits_{\Sigma _{i}}\Phi
_{i}(S_{i})\frac{e^{ikL(S_{i},P_{o})}}{L(S_{i},P_{o})}K(S_{i},P_{o})d\mathbf{l},\label{HF}%
\end{equation}
where $\Phi_{i}(S_{i})$  and $\Phi_{o}(P_{o})$ are, respectively, the incident and output complex fields at the input and output ends, $k=2\pi/\lambda$ is the wavenumber with wavelength $\lambda$, and $d\mathbf{l}$ denotes the 
primary wave source on the incident end $\Sigma _{i}$,
which now is essentially 1D. 
Here, $K=\cos\theta$ is the obliquity factor with $\theta$ being the angle between the propagation axis and
the geodesic connecting any input point $S_{i}$ on $\Sigma _{i}$ and the output point $P_{o}$ on $\Sigma _{o}$, as is shown in the inset of Fig. \ref{figure1}. In Fig. \ref{figure2}(b), it shows that this obliquity factor can be reasonably taken as $1$ when the propagation distance $d$ is one order of magnitude larger than the transverse dimension. 
Here, we emphasize again that both the incident and output ends, $\Sigma _{i}$ and $\Sigma _{o}$, should be taken along the geodesics which are locally orthogonal to the propagation optical axis, thus the points on $\Sigma _{i}$ and $\Sigma _{o}$ are $S_{i}=(x,y,z)$ and $P_{o}=(x^{\prime},y^{\prime},z^{\prime})$, obeying  Eqs. (\ref{geodesic1}) and (\ref{geodesic2}). The function $L(S_{i},P_{o})$ represents the eikonal function which is essentially the length of the geodesic connecting $S_{i}$ on $\Sigma _{i}$
and $P_{o}$ on $\Sigma _{o}$, obtained by\ $L=\int ds$. Technically, except for few special surfaces, the analytical expression of $L$ is scarcely available, leading to the difficulty in the integral in Eq. (\ref{HF}). Therefore in practice, for a certain point $P_{o}$ on $\Sigma _{o}$, we also calculate the geodesic lengths directly, according to Eqs. (\ref{geodesic1}) and (\ref{geodesic2}), connecting $P_{o}$ and hundreds of discrete
points $S_{i}$ on $\Sigma _{i}$, which are deviated from the optical propagation axis as shown in Fig. \ref{figure1}. Thus one can obtain the exact field distribution of light along $\Sigma _{o}$.

In practice, the rotational symmetry exists extensively in many celestial systems.
Here we consider a special family of curved surfaces with rotational symmetry, universally known as SORs, whose metrics can be generally expressed as $ds^{2}=\left[  1+\left(  \frac{dH}{dr}\right) ^{2}\right]  dr^{2} +r^{2}d\varphi^{2}$ in a polar coordinate system for convenience. Thanks to the
orthogonality of the coordinates as well as the rotational symmetry, now one is
able to solve Eqs. (\ref{geodesic1}) and (\ref{geodesic2}) analytically (for mathematical details, see Appendix A),%
\begin{equation}
d\varphi=\pm\frac{\kappa}{r\sqrt{r^{2}-\kappa^{2}}}\sqrt{1+\left(  \frac
{dH}{dr}\right)  ^{2}}dr, \label{geodesicappen}%
\end{equation}
where $\kappa=r_{\text{initial}}^{2}\left(  \frac{d\varphi}{ds}\right)
_{\text{initial}}$ is a constant determined by initial conditions, and the sign of $\pm$
corresponds to two different trajectories according to the sign of the initial
condition $\frac{dr}{ds}$. \ \ \

\subsection{Light propagation on Flamm's paraboloids}
Next we will use the above approach to consider the light propagation on a specific SOR, the Flamm's
paraboloid, as is shown in Fig. \ref{figure2}(a). This interesting surface reveals the spatial curvature in the
vicinity of a Schwarzschild black hole \cite{Throne1973}. As is known, the
gravitational field outside the Schwarzschild radius $r_{s}$ of an uncharged
irrotational spherical mass is described by the Schwarzschild metric
\begin{align}
ds^{2}=& -\left(  1-\frac{r_{s}}{r}\right)  c^{2}dt^{2}+\left(  1-\frac{r_{s}%
}{r}\right)  ^{-1}dr^{2} \nonumber \\
& +r^{2}d\psi^{2}+r^{2}\sin^{2}\psi d\varphi
^{2}.
\label{schwarzschild}%
\end{align}

With its spherical symmetry, the equatorial slice is taken (i.e., $\psi
=\pi/2$) without loss of generality, and the remnant of spatial part of Eq.
(\ref{schwarzschild}) establishes a SOR\ with $H(r)=\pm2\sqrt{r_{s}(r-r_{s})}$ for $r>r_{s}$.
As is shown in Fig. \ref{figure2}(a), first we consider the propagation of a well-collimated Gaussian wavepacket launched along four different geodesics. The Gaussian profile is well maintained at an arbitrary output end when the beam is a little bit away from the event horizon of the black hole. For example, when the beam propagates along the curve (ii) (the purple curve), the beam width changes little from A to B$_{1}$ since the light is far away from the black hole ($r>>r_{s}$), while it increases fast as the light approaches the black hole, such as near the points B$_{2}$ and B$_{3}$. The intensity profiles at these points are shown in the right side of Fig. \ref{figure2}(a). In this situation, we can define the beam width $\sigma$ of light as the full width at
its half-maximum of the intensity profile and track its evolution along the propagation.
The changes of the beam width along the red and green geodesics in Fig. \ref{figure2}(a) are demonstrated in Figs. \ref{figure2}(c) and (d), respectively,
along with their intensity evolution in the corresponding inset figures.
The propagation distance $d$ along the propagation axis is normalized by the Rayleigh distance $z_{r}$ of light in flat space.
It is seen that the light beams in both of these cases diverge rapidly even within a short distance compared to the cases in flat space. This tells us that
the divergence of light beams on such curved surface is greatly amplified due to the large spatial curvature generated from the strong gravitational field near the black hole.
The diverging nature of light beams on Flamm's paraboloid is further revealed in Figs. \ref{figure2}(e) and \ref{figure2}(f), where the variation of $\sigma(d)$
at a certain output end versus the initial beam width $\sigma_{0}$ along the four geodesics in Fig. \ref{figure2}(a) are illustrated.
We can see that the Gaussian beams with small $\sigma_{0}$ are subject to strong divergence, under the same $d$, due to the experience of the strong spatial curvature, and the turning points in these four case are slightly different. These effects are actually related to the strength of negative spatial curvature \cite{Batz2008,XuPRA}.

\begin{figure*}[btp]
\centering
\includegraphics[width=13.5cm]{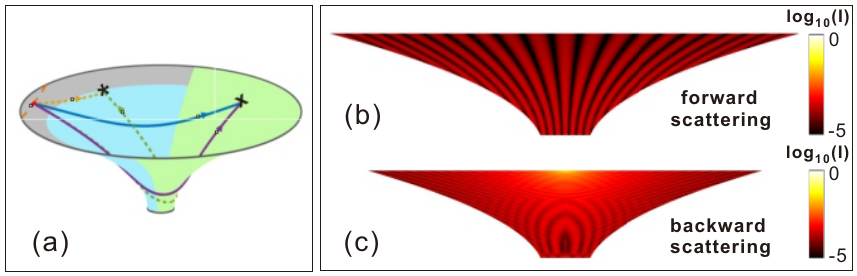}\caption{(a) Sketch of two geodesics from a certain point (red color) on the input to an arbitrary point (black cross or black asterisk) on surface. (c-d) Intensity distributions (denoted by the common logarithm of intensity $I=\left| \Phi_{o} \right|^{2}$) in the green and blue regions of (a) are shown in (b) and (c), respectively, with magnitude being normalized by the maximal intensity of the incident light. Note that the intensity distribution in the gray region of (a) is not shown in (c) in case the intense field near the input end obscures the remaining region in (c). Other parameters are $r_{s}=15$ mm, $\lambda=1.5$ cm, $r(\Sigma_{i})=150$ mm, $\sigma_{0}=1$ mm.}%
\label{figure3}%
\end{figure*}

Now we consider the propagation of a very narrow light beam with a large divergent angle launched directly towards the black hole, along the geodesic (iv) in Fig. \ref{figure2}(a). This is usually similar to the situation that a point-like light source is far away from a black hole. The beam is so divergent that instead of being entirely captured, only the central portion is absorbed by the event horizon, while the periphery grazes the black hole and escapes. As a result, signals can be detected at the opposite side of the light source (i.e., the forward scattering), or even be detected at the same side of the light source (i.e., the backward scattering).
In practice, we calculate the distribution of the light field on the entire surface of Flamm's paraboloid, rather than taking a specific far-field output end as we did above. Interestingly, there are always two possible geodesics from a certain point on the input interface to an arbitrary point on such Flamm's paraboloid surface of the black hole. As an example, in Fig. \ref{figure3}(a), we can see that light rays can travel clockwise and anticlockwise along two different geodesics, respectively, to reach the cross and/or asterisk points in the forward and/or backward directions.
The interplay of these two branches of light rays may lead to
interference fringes on the forward hemi-surface, as is clearly revealed in
Fig. \ref{figure3}(b). Such behaviors are similar to the interference characteristics shown in a recent work by Nambu et al. \cite{Nambu2019}. Furthermore, some portions of light rays may even return as the backward scattering, via the backward geodesics after circling around the black hole, to the vicinity of the incident source, resulting in
complicated interference patterns in Fig. \ref{figure3}(c). We believe such forward scattering and
backward scattering phenomena, in spite of their feebleness, could possibly be used to obtain the structural information of black holes and detect the gravitational effects of invisible and small black holes. More detailed properties of the forward and backward scattering effects for light near a black hole will be further explored elsewhere. 
Indeed, the interaction between two
branches of rays also exists in the collimated light beams investigated in the above, yet one branch dominates and consequently no interference can be
observed (see Appendix B).

\subsection{Deciphering the leftover singularity puzzle}
\begin{figure}[htb]
\centering
\includegraphics[width=7.7cm]{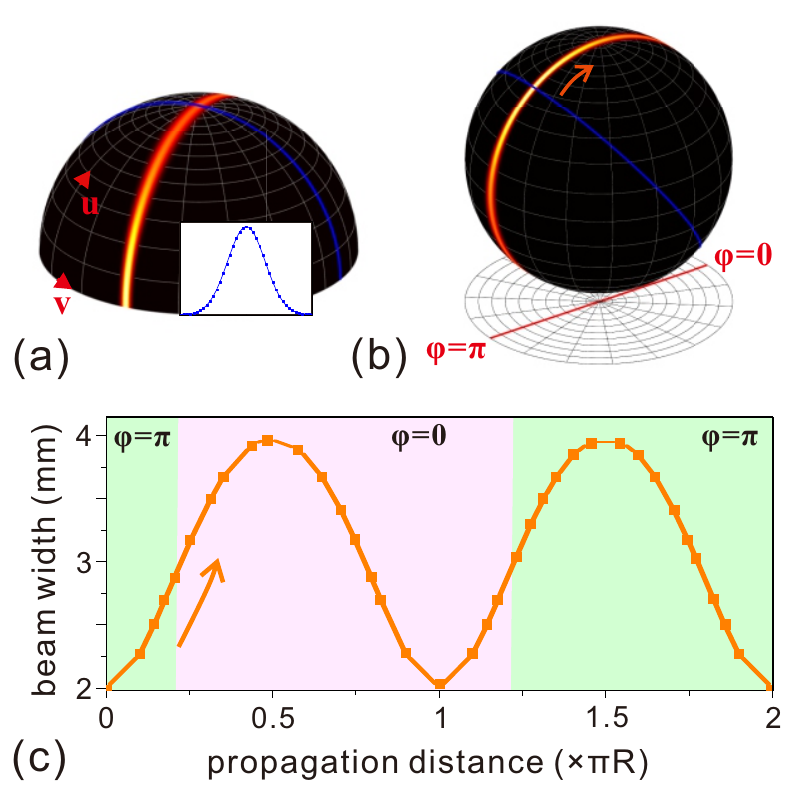}
\caption{(a) The evolution of a light beam along a meridian of a hemisphere, with radius $R$, from the equator. The intensity profile of the light beam at the north pole along the geodesic (the blue line) is shown in the inset. (b) The intensity evolution of a light beam circling around a spherical surface along the direction (denoted by the orange arrow), starting from the geodesic (denoted by the blue solid line), which starts from $r(\Sigma_{i})=0.6R$, $\varphi(\Sigma_{i})=\pi$ on the bottom projected plane. (c) 
Changes of beam width along the propagation on the spherical surface of (b). Other parameters are $R=50$ mm, $\lambda=5\times10^{-4}$ m, $\sigma_{0}=2$ mm.}%
\label{figure4}%
\end{figure}

In Ref. \cite{Batz2008}, an ingenious expression about the propagation of a light beam on the special SORs with constant Gaussian curvature is analytically given by solving the covariant Helmholtz equation in the longitude-latitude coordinate system under paraxial approximation. Take a hemispherical surface in Fig. \ref{figure4}(a) as example, the longitudinal coordinate $u$ along the longitudinal arc direction has the range of $u\in\left[ 0,\pi R/2\right]$ with
$R$ being the radius of hemisphere, while the latitudinal coordinate $v$, being the rotational angle, is within the range $\left[ 0,2\pi \right]$. The arclength of latitudes is thus $2\pi R\cos\left(u/R\right)$, which varies with $u$ and vanishes at the north pole $u=\pi R/2$, resulting in a mathematical singularity. The intensity evolution of a
Gaussian beam propagating along the meridian and starting from the equator follows
\begin{align}
I(u,v)  & =\frac{1}{\cos\left(  \frac{u}{R}\right)  \left[  1+\left.
R^{2}\tan^{2}\left(  \frac{u}{R}\right)  \right/  z_{r}^{2}\right]  ^{\frac
{1}{2}}}\nonumber\\
& \times\exp\left[  -\frac{kz_{r}R^{2}v^{2}}{R^{2}\tan^{2}\left(  \frac{u}%
{R}\right)  +z_{r}^{2}}\right]  ,
\end{align}
where $z_{r}=k\sigma_{0}^{2}/2$ is the Rayleigh distance in flat space, and accordingly the beam width obeys%
\begin{equation}
\sigma(u)=\sigma_{0}\sqrt{\cos^{2}\left(  \frac{u}{R}\right)
+\frac{R^{2}}{z_{r}^{2}}\sin^{2}\left(  \frac{u}{R}\right)
}\label{sigmas}%
\end{equation}
where $\sigma_{0}$ is the initial beam width (for the detailed derivation, see Appendix C). Apparently the intensity is infinite at $u=\pi R/2$,
yet the beam width is not vanishing at the singularity. Such singularity won't occur when the light beam propagates along the equator, which, however, should be exactly identical to the propagation along a meridian, owing to the perfect
symmetry of a sphere. Clearly, such singular behaviors along a meridian are artificial. 

Fundamentally, these puzzles are consequences of mischoice of the far-field output interface and can be well solved by the method
mentioned in this work. In the longitude-latitude coordinate system, both the
incident and far-field interfaces are supposed to be taken along the
latitudinal lines, which are actually not geodesics (except the equator). In practice,
when an observer stands on the curved surface, the coordinates should be
taken locally along the two orthogonal geodesics. In this situation, the propagation of a light beam is
illustrated in Fig. \ref{figure4}(a). Especially, on
the north pole, the geodesic perpendicular to the propagation axis is marked
by the blue bold line, thus the north pole is naturally a regular point instead
of an artificial singularity. Furthermore, we also inspect the propagation of a light beam
circling a sphere in Fig. \ref{figure4}(b), and find that the evolution of
beam width oscillates after each half circumference on the spherical surface. This
interesting property was discovered under the condition
that the propagation is along the equator via the coordinate transformation \cite{Batz2008}. By our method,
this periodic oscillation of the beam width can also be obtained when the incident Gaussian beam starts from arbitrary positions on
sphere, which is unavailable in the previous method. 
These results not only
solve the artificial singularity enigma, but are also more pragmatic in real experiments,
since it is more appropriate to take an arbitrary geodesic as the incident interface when a laser beam is coupled onto a curved surface.
\subsection{Possible experimental schemes}
Our theory can be experimentally implemented both macroscopically and microscopically. In pioneering experiments, constraining light propagation on curved surfaces was realized either by total internal reflection in curved crown glass \cite{Schultheiss2010, Schultheiss2015, Patsyk2018} or by a thin liquid waveguide covered on a 3D solid object \cite{Schultheiss2010, Schultheiss2015}. The 3D objects with prescribed shapes can be fabricated by state-of-the-art technologies, such as high precision diamond turning for macroscopic structures \cite{Schultheiss2010} and Nanoscribe 3D laser lithography technique \cite{Bekenstein2017, Maruo1997} in nanometric scale. Very recently, an intriguing work \cite{Segev2020} demonstrates light propagation on thin curved soap membranes (see its supplementary video 3). This scheme provides a novel promising platform, especially when the varying thickness of membrane, acting as effective refractive index, could be an extra dimension for modulation. Moreover, it is proved that a curved surface of revolution is equivalent to a plane with azumuthally symmetric distribution of refractive index \cite{arxiv2020}. Therefore, an alternative pathway is to fabricate the predesigned refractive index profile on a planer surface, by, for example, a microsphere-embedded variable-thickness polymethyl methacrylate waveguide \cite{Sheng2013, Sheng2015}, or through the optically-induced giant Kerr effect in liquid crystal \cite{Khoo2011}, with its landscape being provided by a spatial light modulator working in reflection.

\section{CONCLUSION}
In conclusion, we develop heuristics to study the propagation of light in 2D curved space, founded on Huygens-Fresnel principle.
This method is feasible when the direction of light propagation is along arbitrary geodesics on any curved surfaces.
By this method, we study the behaviors of light beams on a Flamm's paraboloid, which is the 2D correspondence of spatial curvature
outside a Schwarzschild black hole. We investigate the evolution of Gaussian wavepackets propagating along different geodesics,
and reveal the diverging nature of light on such curved surface. We also illustrate the interference patterns induced by highly divergent light sources.
Finally we point out that this method works out the remaining puzzles about the coordinate singularities in the previous theory.
Our work provides a powerful tool and refreshing insights which greatly broaden the possibilities of investigations about light propagation in curved space. Exotic geodesics \cite{Besse1978} could be utilized to realize lights with special properties. With the help of the proposed method, further studies could be extended to other optical effects, such as spectral properties \cite{XuPRA, XuOE}, phase information \cite{XuNJP}, Hanbury Brown and Twiss effect \cite{Schultheiss2015}, geodesic lens \cite{glens2020}, Talbot effect \cite{Hall2021} and acceleration radiation \cite{Scully2018}. Moreover, the investigations about optics on Flamm's paraboloids open up a new perspective on the radiations in the vicinity of Schwarzschild black holes, and contribute to the interdisciplinary explorations of cosmology and optics.

\section*{APPENDIX A: DERIVATION OF EQ. (5)}
When written in polar coordinates, the metric of general surfaces of
revolution is%
\begin{equation}
ds^{2}=\left[ 1+\left( \frac{dH}{dr}\right) ^{2}\right] dr^{2}+r^{2}d\varphi
^{2}.  \label{metriccc}
\end{equation}%
Therefore, the geodesic equations take the form
\begin{equation}
\frac{d^{2}r}{ds^{2}}+\frac{\frac{dH}{dr}}{1+\left( \frac{dH}{dr}\right) ^{2}%
}\frac{d^{2}H}{dr^{2}}\left( \frac{dr}{ds}\right) ^{2}-\frac{r}{1+\left(
\frac{dH}{dr}\right) ^{2}}\left( \frac{d\varphi }{ds}\right) ^{2}=0,
\label{rgeodesic}
\end{equation}%
\begin{equation}
\frac{d^{2}\varphi }{ds^{2}}+\frac{2}{r}\frac{dr}{ds}\frac{d\varphi }{ds}=0.
\label{faigeodesic}
\end{equation}%
Eq. (\ref{faigeodesic}) can be solved as%
\begin{equation}
\frac{d\varphi }{ds}=\frac{\kappa }{r^{2}},  \label{dfaids}
\end{equation}%
where $\kappa =r_{\text{initial}}^{2}\left( \frac{d\varphi }{ds}\right) _{%
\text{initial}}$ is an integration constant whose value is determined by
initial conditions of the geodesic. Besides, since Eq. (\ref{metriccc}) is
basically a spatial metric (i.e., $ds>0$), one can divide both sides by $%
ds^{2}$, and readily have
\begin{equation}
\frac{dr}{ds}=\pm \frac{1}{\sqrt{1+\left( \frac{dH}{dr}\right) ^{2}}}\sqrt{%
1-\frac{\kappa ^{2}}{r^{2}}}.  \label{drds}
\end{equation}%
Here $+(-)$ is taken when $\left( \frac{dr}{ds}\right) _{\text{initial}%
}>\left( <\right) 0$, therefore $\pm $ corresponds to two different
geodesics. Eq. (\ref{drds}) can also be obtained by solving Eq. (\ref{rgeodesic}) with Eq. (\ref{dfaids}), by the method of constant variation.
At last, after dividing Eq. (\ref{dfaids}) by Eq. (\ref{drds}), we have Eq.
(5) in the main text,%
\begin{equation}
\frac{d\varphi }{dr}=\pm \frac{\kappa }{r^{2}\sqrt{1-\frac{\kappa ^{2}%
}{r^{2}}}}\sqrt{1+\left( \frac{dH}{dr}\right) ^{2}}.
\end{equation}

\section*{APPENDIX B: SUPERPOSITION OF CLOCKWISE (CW) AND ANTICLOCKWISE (ACW) GEODESICS}
\begin{figure*}[htb]
\centering
\includegraphics[width=13cm]{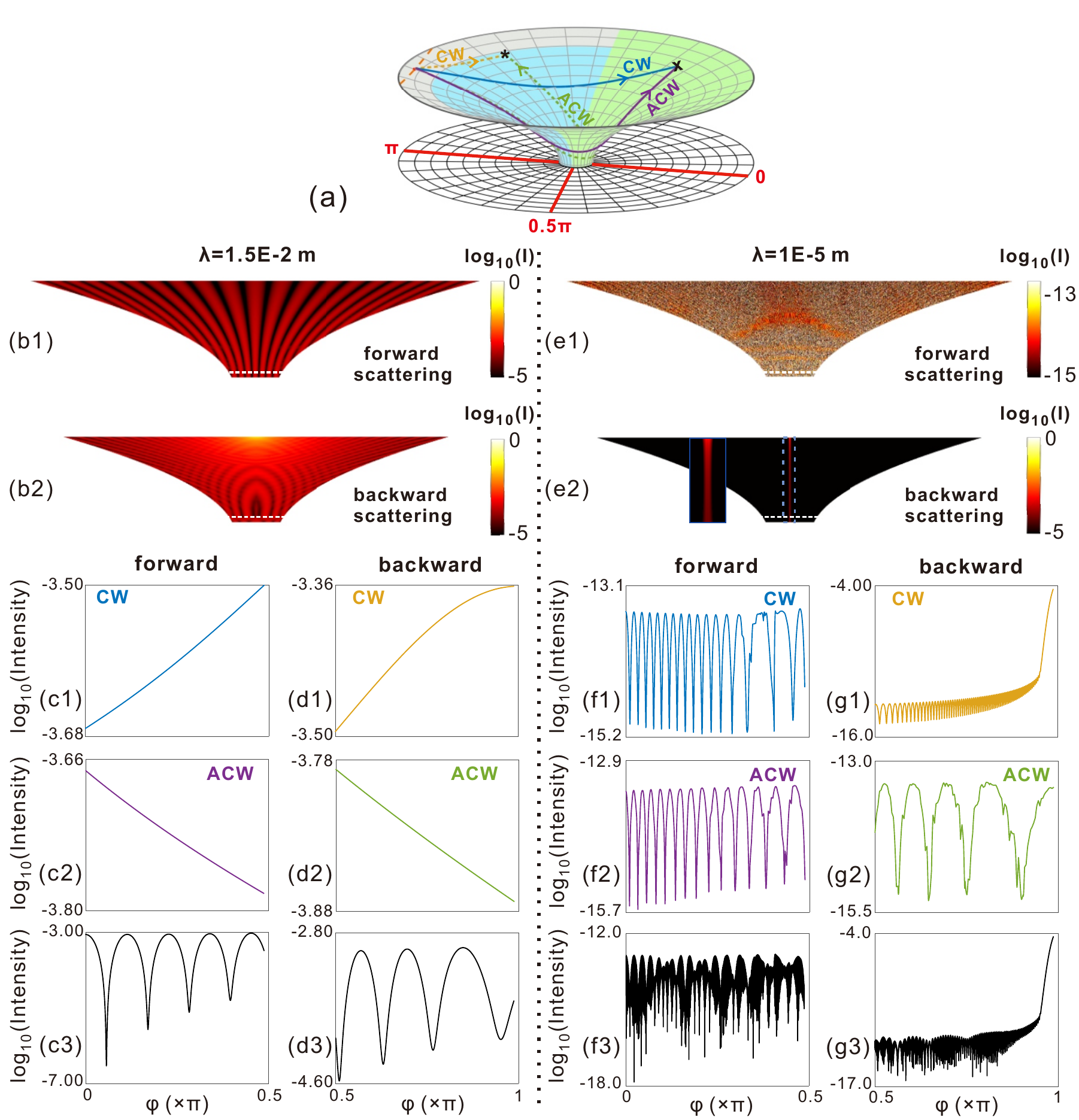}\caption{(a) Sketch of a Flamm's paraboloid. (b1)(b2) Intensity distribution on the incident (b1) and opposite (b2)
hemi-surface with wavelength $\lambda =1.5\times 10^{-2}$ m. These two subfigures are same as Figs. \ref{figure4}(b) and (c). (c1)(c2)(d1)(d2) Intensity of the
field induced exclusively by CW (c1)(d1) and ACW (c2)(d2) geodesics along a certain latitude [white dashed lines in (b1) and (b2)] on the opposite
hemi-surface (c1)(c2) and incident hemi-surface (d1)(d2). Due to symmetry,
only a quarter of surface is plotted, with $\varphi $ labeled in (a).
(c3)(d3) Intensity of the field after superimposing the complex amplitudes
induced by CW and ACW geodesics. (e1)(e2) Intensity distribution of a light
beam with same parameters except for wavelength $\lambda =1\times 10^{-5}$
m, which is therfore less divergent. Inset of (e2) is the zoom of the area
in the blue dashed frame. (f1)(f2)(f3)(g1)(g2)(g3) Intensity profiles along white dashed latitudes in (e1) and (e2).
}%
\label{figure5}%
\end{figure*}
In Fig. \ref{figure3}(a), we have illustrated that there are always two geodesics
from the point on the input end to an arbitrary point on Flamm's
paraboloid, one traveling in CW direction (i.e., along the blue solid line to the cross point and along the yellow dashed line to the asterisk point) and another in ACW direction (i.e., along the purple solid line to the cross point and along the
green dashed line to the asterisk point). The interaction of these two branches of geodesics leads
to interference patterns on the entire surface for highly divergent light.
In essence, the existence of these two branches of geodesics is a geometrical
property of surface \textit{per se}, regardless of the incident light beams. Therefore, theoretically, the intensity distribution of a collimated light
beam (for example, the ones in Fig. \ref{figure2}) is also a result of the superposition
of CW and ACW geodesics. In Figs. \ref{figure5}(e1) and (e2), we illustrate a light
beam with same initial beam width but much smaller wavelength (that is, much
longer Rayleigh distance and thus more collimated) as that in Fig. \ref{figure4} as well
as Figs. \ref{figure5}(b1) and (b2). It is clearly seen that at the incident-side half hemi-surface
which contains the incident field, the Gaussian profile of intensity
almost remains intact during propagation and the light beam barely diverges,
with no interference patterns observed. While at another half hemi-surface
in the forward direction, the intensity is extremely feeble,
indicating that no signal exists up there, since the black hole blocks or absorbs almost all light energy. These results accord with that shown in
Fig. \ref{figure2}, where only one geodesic is taken into consideration.
As a matter of fact, when we further inspect the contributions of CW and ACW
geodesics in Figs. \ref{figure5}(g1)-(g3), we find out that for collimated light beams,
the field induced by one branch (ACW in this case) is way smaller than another,
or in other words, the contribution from CW geodesic dominates and thus no
interference phenomenon occurs. In contrast, for highly diverging light beams,
the fields induced by both CW and ACW geodesics are at a comparable level
as is shown in Figs. \ref{figure5}(c1-c2) and Figs. \ref{figure5}(d1-d2), and therefore
interference fringes present in Figs. \ref{figure5}(c3) and (d3).

\section*{APPENDIX C: DERIVATION OF EQS. (7) AND (8)}
The electric field of light on curveds surface obeys the covariant Helmholtz
equation%
\begin{equation}
\frac{1}{\sqrt{g}}\partial _{i}\left( \sqrt{g}g^{ij}\partial _{j}\Phi
\right) +k^{2}\Phi =0,  \label{we1}
\end{equation}%
where $k$ is the wave number, $g^{ij}$ is the element of the
inverse matrix of metric tensor $\mathbf{g}$, and $g$ is the determinant of $%
g_{ij}$. To construct a spherical surface in Fig. \ref{figure4}, we take the
longitudinal arc length as the longitudinal coordinate $u$, and the rotational angle
as the latitudinal coordinate $v$, the spherical surface can thus be defined by $%
r_{\text{ROR}}(u)=R\cos \left( \frac{u}{R}\right) $, with $r_{\text{ROR}}$
being radius of revolution and $R$ being the radius of sphere. With the
position vector $\mathbf{\rho }=\left[ R\cos \left( \frac{u}{R}\right) \cos
v,R\cos \left( \frac{u}{R}\right) \sin v,R\sin \left( \frac{u}{R}\right) %
\right] $, one is able to calculate the metric by $g_{ij}=\frac{\partial
\mathbf{\rho }}{\partial x^{i}}\cdot \frac{\partial \mathbf{\rho }}{\partial
x^{j}}$, where $i,j=1,2$ and $x^{1}=u$, $x^{2}=v$. Therefore, the wave equation,
Eq. (\ref{we1}), on spherical surfaces is expressed as%
\begin{equation}
\frac{\partial ^{2}\Phi }{\partial u^{2}}-\frac{\tan \left( \frac{u}{R}%
\right) }{R}\frac{\partial \Phi }{\partial u}+\frac{1}{R^{2}\cos ^{2}\left(
\frac{u}{R}\right) }\frac{\partial ^{2}\Phi }{\partial v^{2}}+k^{2}\Phi =0,
\label{we2}
\end{equation}%
with metric $g_{11}=1$, $g_{22}=R^{2}\cos ^{2}\left( \frac{u}{R}\right) $.

Taking the ansatz $\Phi (u,v)=\left[ \cos \left( \frac{u}{R}\right) \right]
^{-\frac{1}{2}}\Psi (u,v)$, after tedious mathematics, one has
\begin{equation}
\frac{\partial ^{2}\Psi }{\partial u^{2}}+\frac{1}{R^{2}\cos ^{2}\left(
\frac{u}{R}\right) }\frac{\partial ^{2}\Psi }{\partial v^{2}}+k_{\text{eff}%
}^{2}\Psi =0,  \label{we3}
\end{equation}%
with $k_{\text{eff}}^{2}=k^{2}+\Delta $ and $\Delta =\frac{1}{4R^{2}}\left[
1+\cos ^{-2}\left( \frac{u}{R}\right) \right] $.
We assume $\Psi (u,v)=\Xi (u,v)\exp \left( ik_{\text{eff}}u\right) $, with
paraxial approximation
\begin{equation}
\frac{\partial ^{2}\Xi }{\partial u^{2}}\ll 2k\frac{\partial \Xi }{\partial u%
}\text{,}  \label{paraxial}
\end{equation}%
being carried out, and subsequently let $\Xi (u,v)=\phi (u,v)\exp \left[
\frac{i}{2k}\int \Delta (u^{\prime })du^{\prime }\right] $, we eventually
reach
\begin{equation}
2ik\frac{\partial \phi }{\partial u}+\frac{1}{R^{2}\cos ^{2}\left( \frac{u}{R%
}\right) }\frac{\partial ^{2}\phi }{\partial v^{2}}=0.  \label{we4}
\end{equation}

For Eq. (\ref{we4}), suppose the solution takes the form
\begin{equation}
\phi (u,v)=\exp \left[ i\alpha (u)+\frac{ikv^{2}}{2\beta (u)}\right] ,
\label{fai}
\end{equation}%
where $\alpha (u)$ and $\beta (u)$ are functions which satisfy%
\begin{equation}
-2k\frac{\partial \alpha (u)}{\partial u}+\frac{1}{R^{2}\cos ^{2}\left(
\frac{u}{R}\right) }\frac{ik}{\beta (u)}=0,  \label{1}
\end{equation}%
\begin{equation}
\frac{\partial \beta (u)}{\partial u}-\frac{1}{R^{2}\cos ^{2}\left( \frac{u}{%
R}\right) }=0.  \label{2}
\end{equation}%
Here we suppose that the light launches at the equator and is in a Gaussian
profile with the initial beam width $\sigma _{0}$, i.e., $\phi (u=0,v)=\exp %
\left[ -\frac{R^{2}v^{2}}{\sigma _{0}^{2}}\right] $, Eqs. (\ref{1}) and (\ref%
{2}) can be readily solved as%
\begin{equation}
\beta (u)=\frac{1}{R}\tan \left( \frac{u}{R}\right) -\frac{i z_{r} }{R^{2}%
},  \label{beta}
\end{equation}%
\begin{equation}
\alpha (u)=i\left\{ \frac{\ln \left[ \frac{z_{r} ^{2}+R^{2}\tan
^{2}\left( \frac{u}{R}\right)}{z_{r} ^{2}} \right]}{4} +\frac{i}{2}\arctan \left[ \frac{%
R\tan \left( \frac{u}{R}\right) }{z_{r} }\right] \right\},
\label{alpha}
\end{equation}%
with $z_{r} =k\sigma_{0}^2/2$. Therefore we eventually have the solution of Eq. (\ref{we1})
\begin{eqnarray}
\Phi (u,v) &=&\left[ \cos \left( \frac{u}{R}\right) \right] ^{-\frac{1}{2}%
}\left[ \frac{z_{r} ^{2}+R^{2}\tan ^{2}\left( \frac{u}{R}\right)}{z_{r} ^{2}} \right] ^{%
-\frac{1}{4}}  \nonumber \\
&&\times \exp \left\{ -\frac{kz_{r} R^{2}v^{2}}{2\left[ z_{r}
^{2}+R^{2}\tan ^{2}\left( \frac{u}{R}\right) \right] }\right\}   \nonumber \\
&&\times \exp \left\{ -\frac{i}{2}\arctan \left[ \frac{R\tan \left( \frac{u%
}{R}\right) }{z_{r} }\right] \right\}  \nonumber
\\
&&\times \exp \left[ \frac{ikR^{2}v^{2}}{2}\frac{R\tan \left( \frac{u}{R}%
\right) }{R^{2}\tan ^{2}\left( \frac{u}{R}\right) +z_{r} ^{2}}\right]  \nonumber \\
&&\times \exp \left[ \frac{i}{2k}\int \Delta (u^{\prime })du^{\prime }\right] \exp \left[ iku\right],
\label{field}
\end{eqnarray}%
from which we further have the intensity%
\begin{eqnarray}
I(u,v) &=&\Phi ^{\ast }(u,v)\Phi (u,v)  \nonumber \\
&=&\left[ \cos \left( \frac{u}{R}\right) \right] ^{-1}\left[ 1+\frac{R^{2}\tan ^{2}\left( \frac{u}{R}\right)}{z_{r}^{2}} \right] ^{-\frac{1}{2}}
\nonumber \\
&&\times \exp \left[ -\frac{kz_{r} R^{2}v^{2}}{z_{r} ^{2}+R^{2}\tan
^{2}\left( \frac{u}{R}\right) }\right]   \label{intensity}
\end{eqnarray}%
and beam width%
\begin{equation}
\sigma (u)=\sigma _{0}\sqrt{\cos ^{2}\left( \frac{u}{R}\right) +\frac{R^{2}}{%
z_{r} ^{2}}\sin ^{2}\left( \frac{u}{R}\right) }.  \label{sigmas}
\end{equation}

\begin{acknowledgments}
 Zhejiang Provincial Natural Science Foundation of China (No. LD18A040001), the National Natural Science Foundation of China (NSFC) (No. 11974309 and 11674284), and National Key Research and Development Program of China (No. 2017YFA0304202).
\end{acknowledgments}

\end{document}